\documentclass[twocolumn,shownopacs,showkeys,preprintnumbers,amsmath,amssymb,unsortedaddress]{revtex4-1} 

\usepackage{graphicx,color}

\usepackage{dcolumn}
\usepackage{bm,multirow}

\begin{document}

\title{Topological constraints in eukaryotic genomes and how they can be exploited to improve spatial models of chromosomes}

\author{Angelo Rosa}
\email{anrosa@sissa.it}
\affiliation{
SISSA (Scuola Internazionale Superiore di Studi Avanzati), Via Bonomea 265, 34136 Trieste (Italy)
}
\author{Marco Di Stefano}
\email{marco.distefano@cnag.crg.eu}
\affiliation{
CNAG-CRG (Centre Nacional d'An\`alisi Gen\`omica-Centre de Regulaci\'o Gen\`omica), Carrer de Baldiri Reixac, 4, 08028 Barcelona, Spain
}
\author{Cristian Micheletti}
\email{michelet@sissa.it}
\affiliation{
SISSA (Scuola Internazionale Superiore di Studi Avanzati), Via Bonomea 265, 34136 Trieste (Italy)
}


\begin{abstract}
Several orders of magnitude typically separate the
contour length of eukaryotic chromosomes and the size of
the nucleus where they are confined.
The ensuing topological constraints can slow down the
relaxation dynamics of genomic filaments to the point that mammalian
chromosomes are never in equilibrium over a cell's lifetime.
In this opinion article, we revisit these out-of-equilibrium effects and discuss how their
inclusion in physical models can enhance the spatial reconstructions of interphase eukaryotic genomes from
phenomenological constraints collected during interphase.
\end{abstract}

\maketitle

\section{Introduction}

From viruses to eukaryotes, genomic DNA filaments are confined in spaces of linear dimension much smaller than their contour lengths.
In bacteriophages, the $\mu$m-long genome is stored in $\approx 50$nm-wide viral capsids
and the corresponding packing density is so high that viral DNA filaments that have little chance to be entangled when in solution (knotting probability $<3\%$) become almost certainly knotted ($>95\%$ probability) once confined inside capsids~\citep{Rybenkov_et_al_PNAS_1993,arsuaga2002,Marenduzzo_et_al_PNAS_2009,Marenduzzo_et_al_2010_JoP_Cond_Matt}.
In humans, instead, the various cm-long chromosomes that make up the genome are kept inside $\approx 10\mu$m-wide nuclei~\citep{alberts}.
Despite the major change of scale with respect to viruses, the volume fraction occupied by this eukaryotic genome is still large, about $10\%$~\citep{RosaEveraersPlos2008}.

These considerations pose several conundrums:
How can chromosomal DNA be at the same time packed and yet accessible to the regulatory and transcriptional machineries?
What is its typical degree of genomic entanglement and how much does it interfere with DNA transactions?
To what extent are these aspects shaped by general passive physical mechanisms versus active ones, {\it e.g.} involving topoisomerase enzymes?

\section{Intra- and inter-chromosome architecture}

\subsection{Phenomenology}

Addressing these questions has proved challenging because of the wide range of length and time scales involved in genome architecture.
Classical experimental tools provide details of chromosome architecture at two opposite scales~\citep{MarcGap}.
At the smallest one ($10-100$nm) X-ray crystallography revealed that DNA achieves local packing by wrapping around histones, while at the largest one ($1-10\mu$m) fluorescence {\it in situ} hybridization (FISH) showed that each chromosome occupies a compact region of the nucleus, termed {\it territory}~\citep{CremerBrosReview2001,CremerBrosReview2010}.

More recently, experimental breakthroughs such as
super-resolution imaging,
electron microscopy tomography plus selective labelling and chromosome conformational capture (Hi-C) techniques have significantly extended our ``multiscale'' knowledge of genome architecture ~\citep{BoettigerZhuangNature2016,BintuZhuangScience2018,Nir2018,OSheaChromEMTScience2017,dekker3c,hic}.

These and other advancements helped establishing various results that foster the present discussion of genomic entanglement.

Regarding inter-chromosome organization we recall that:
\begin{itemize}
\item[(i)] the positioning of chromosome territories correlates significantly with sequence-dependent properties of the underlying DNA (most notably, gene density~\citep{BolzerCremerBrosPlosBiol2005});
\item[(ii)] the intermingling of different chromosomes is minimal and mostly restricted to the boundaries of the territories~\citep{CremerBrosReview2001,BrancoPomboPlosBiol2006}.
\end{itemize}

For intra-chromosome aspects we instead know that:
\begin{itemize}
\item [(iii)] on the scale of a few kilo-basepairs up to about 1 mega-basepair, chromosomes are organized into self-interacting regions, called {\it topologically-associating domains} or TADs \citep{DixonNature2012,NoraNature2012}. On the tens of mega-basepairs scale, chromatin is organized into compartments of varying compactness depending on their functional and epigenomic state~\citep{hic,WangZhuangScience2016}.
\item [(iv)] despite this variability, when averaged over chromosomes and experimental realizations, the mean contact probability of two chromosomal {\it loci} at genomic distance $\ell$ scales approximately as $\langle p_c (\ell) \rangle \sim \ell^{-1}$~\citep{hic}, and the mean square separation scales as $\langle R^2 (\ell) \rangle \sim \ell^{2/3}$~\citep{sachs1,Langowski1999}.
\end{itemize}

\subsection{Relating genomic architecture and relaxation dynamics with polymer physics}

The interpretation of these experimental results has been aided by an intense theoretical and computational activity that demonstrated how salient genomic architecture properties can be reproduced by a broad range of polymer models, and hence are likely governed by general physical mechanisms~\citep{MirnyReview2011,RosaZimmer2014,BiancoNicodemiReview2017,JostRosaVaillantEveraersReview2017,HaddadJostVaillant2017,TianaGiorgettiCOSB2018}.
This applies in particular to the aforementioned properties (i-iv) which can be rationalised as manifestations of the topological constraints that rule the behavior of semi-dilute or dense polymer systems by having a dramatic impact on their relaxation time scales~\citep{DoiEdwards}.

In fact, a solution of initially disentangled chains of contour length $L_c$ can reach the fully-mixed, homogeneous equilibrium state only via reptation, a slow and stochastic slithering-like motion with characteristic time scale equal to $\tau_{\rm rept} \simeq \tau_{e} (L_c / L_e)^3$, where $\tau_{e}$ is a microscopic collision time and $L_e$ is  the typical contour length between entanglement points~\citep{deGennes71,DoiEdwards}.

Thus, based on this fundamental polymer physics result, it was estimated that the characteristic relaxation, or equilibration, time of mammalian chromosomes exceeds 100 years~\citep{RosaEveraersPlos2008}.
The orders-of-magnitude difference between this time scale and the typical duration of the cell cycle ($\approx1$ day) has several implications for genome organization, as it was realised even before Hi-C probing methods became available~\citep{RosaEveraersPlos2008}.
It is clear, in fact, that mammalian chromosomes are never fully relaxed as they undergo the cyclic structural rearrangements from the separate compact rod-like mitotic architecture to the decondensed interphase one~\citep{grosbergEPL1993,RosaEveraersPlos2008}.

\subsection{Implications for (minimal) intra- and inter-chromosome entanglement}

From this standpoint, the emergence of chromosome territories is quantitatively explained as due to the kinetically trapped 
decondensation of the compact mitotic chromatin~\citep{RosaEveraersPlos2008}: 
interphase chromosomes retain the {\it memory} and {\it limited mutual overlap} of the earlier mitotic state,
consistent with
experimental results~\citep{CremerBrosReview2001,BrancoPomboPlosBiol2006,CremerBrosReview2010}.
In addition, the ordered linear organization of the mitotic rods should also inform the intra-chromosomal architecture, making it more local than equilibrated polymers.
This is consistent with the experimental fact that the effective scaling behaviour of the contact probability with the genomic separation $\ell$ in interphase chromosomes has a more local character ($\sim \ell^{-1}$) than the one expected ($\sim \ell^{-3/2}$) for equilibrated polymers~\citep{hic}.
Intuitively, the same ``memory'' mechanism ought to facilitate the subsequent separation of interphase chromosomes and their recondensation upon re-entering the mitotic phase in the cell cycle~\citep{RosaEveraersPlos2008}.

For the present discussion, we stress that these out-of-equilibrium effects should impact not only the architecture but also the physical entanglement of eukaryotic genomes. In fact, mammalian chromosomes should be more unlinked (for the limited inter-chromosomal intermingling) and unknotted (for the enhanced intra-chromosomal local contacts) than at equilibrium.
These heuristic conclusions are supported by various studies showing that the aforementioned scaling relationships obtained by FISH and Hi-C experiments can be ascribed to the topological constraints at play in solutions of unknotted and unlinked polymers~\citep{KhokhlovNechaev85,VettorelPhysToday2009,HalversonSmrekKremerGrosbergReview2014,RosaEveraersPRL2014}.

\subsection{Implications for genomic structural modelling and its improvement}

These considerations appear particularly relevant for the structural modelling of eukaryotic genomes based on phenomenological data, such as spatial proximity constraints, which are typically too sparse~to pin down even coarse-grained models of interphase chromosomes~\citep{hic}.

A key question is whether such structural models should {\em additionally} be informed by the notion that interphase chromosomes must originate and eventually return to the separate and condensed mitotic state.

\begin{figure*}[h!]
\begin{center}
\includegraphics[width=0.9\textwidth]{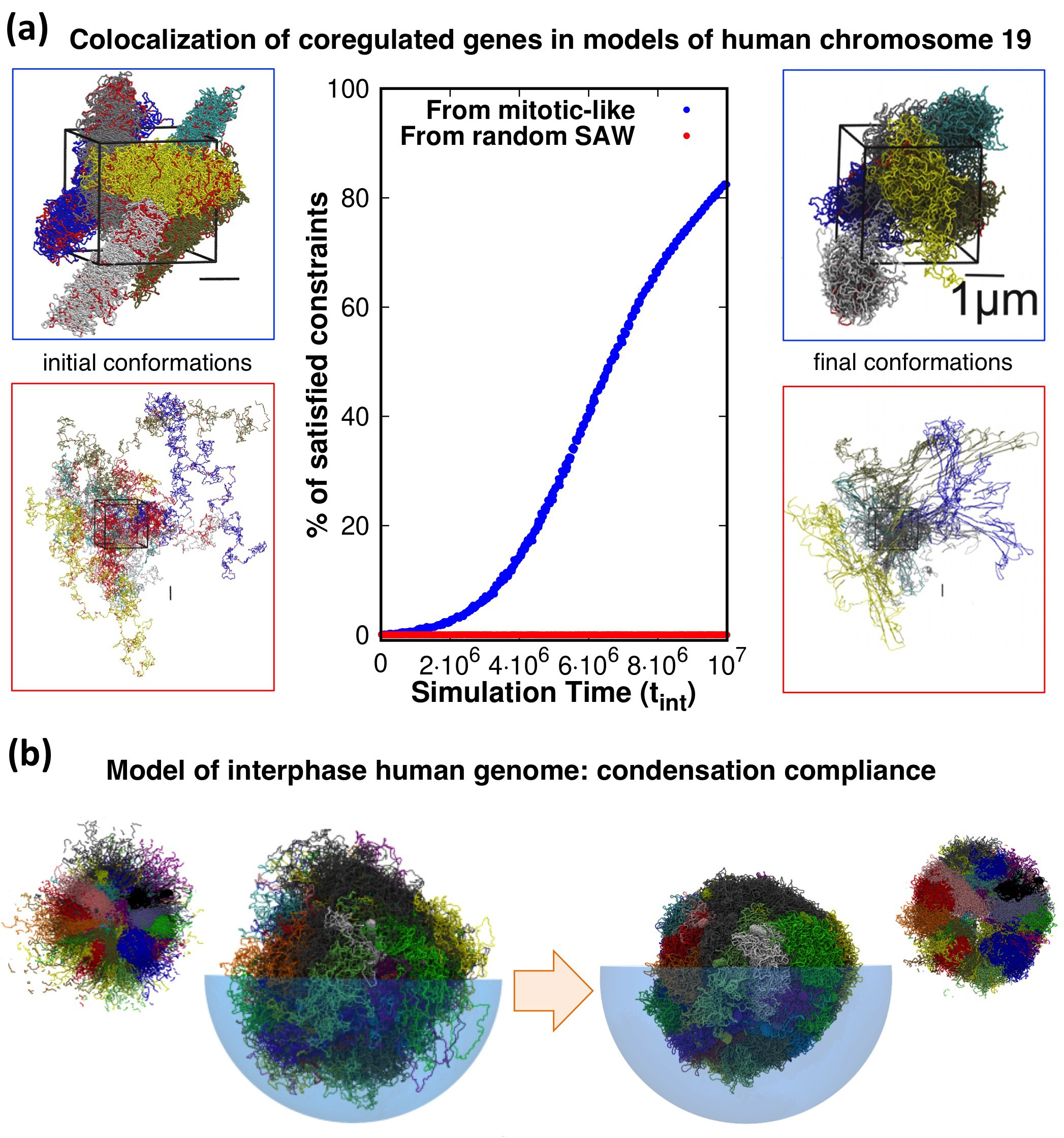}
\end{center}
\caption{(a) Model conformations of human chromosome 19 (six copies, arranged in a periodic simulation box), described as self-avoiding chains of beads were reshaped by steered molecular dynamics (MD) simulations that promoted the colocalization of pairs of {\em loci} that are significantly coregulated. Most ($>$80$\%$) of the coregulated pairs were successfully brought into spatial proximity in simulations that started from relaxed solenoidal mitotic-like arrangements, while virtually no successful colocalization was observed for trajectories started from equilibrated, fully mixed, chromosomal arrangements. Adapted with permission from~\citep{DiStefanoRosa2013}. (b) Model conformations of the entire-human genome, obtained steered-MD colocalization of {\em loci} based on Hi-C data in~\citep{DixonNature2012,Rao2014} could be succesfully condensed with minimal hindrance from intra- or inter-chromosomal constraints, consistently with the expected reconfiguration compliance necessary for the interphase$\to$mitotic transition. The smaller side pictures are cut-through views. Adapted with permission from~\citep{DiStefanoSciRep2016}.}
\label{fig:1}
\end{figure*}

Evidence presented in our earlier work help shed some light on the matter.
With our co-workers, we considered a model system of 6 copies of human chromosome 19 in a cubic simulation box with periodic boundary conditions to explore the connection between coregulation and colocalization of genes~\citep{DiStefanoRosa2013}.
Each copy was initially prepared as a mitotic-like conformation~\citep{RosaEveraersPlos2008}, consisting of a polymer filament forming a solenoidal pattern with rosette-like cross-section featuring chromatin loops of about $50$ kilo-basepairs, see Fig.~\ref{fig:1}a.
We then used a molecular-dynamics steering protocol to bring in proximity pairs of intra-chromosomal {\it loci} that were known to be significantly co-regulated.
Importantly, topological constraints were accounted for by avoiding unphysical chain crossings during the steering process.

Remarkably, and consistently with the gene kissing hypothesis~\citep{Cavalli2007}, we found that most ($>$ 80\%) pairs of significantly coregulated genes could indeed be colocalised in space within the contact range
of 120nm and further showed that this colocalization compliance followed from the presence of gene cliques in the coregulatory network~\citep{DiStefanoRosa2013}.

Conversely, the same protocol applied to the same set of chains but initially prepared as generic self-avoiding random walks failed to give colocalization~\citep{DiStefanoRosa2013}:
physically, this happens because the intra- and inter-chain entanglements present in the original system, by mimicking an ``artificially'' equilibrated set of chromosomes, were too numerous and conflicting to be successfully negotiated on a viable simulation time scale, see Fig.~\ref{fig:1}a.

Further elements come from the genome-wide structural modelling of human chromosomes of~\citet{DiStefanoSciRep2016}. In this study too, the model chromosomes were initially prepared in mitotic-like states and were then steered to bring in proximity those pairs of {\em loci} that corresponded to significantly enhanced entries of two independent Hi-C datasets~\citep{DixonNature2012,Rao2014}. The architecture of the final conformations were, as expected, significantly changed by the steering protocol. Yet, as illustrated in Fig.~\ref{fig:1}b, we verified that each model chromosome could be brought to a condensed compact shape as needed for the interphase-mitotic transition without significant hindrance from intra- or inter-chromosomal topological constraints~\citep{DiStefanoSciRep2016}.

We note that the limitedly-entangled architecture of models of long eukaryotic chromosomes has emerged lately~\citep{DiPierroWolynesPNAS2016} as the consequence of {\it microphase separation} of regions of different chromatin types~\citep{JostCarrivainNAR2014} in a block co-polymer model with pair interactions tuned to reproduce the contact propensities of point (iv).
The point is reinforced by studies on the yeast genome showing that knots and links have a generally low incidence especially in comparison to equivalent systems of equilibrated chains~\citep{duan2010,segal2014,arsuaga2019}.
Finally, besides the indication from structural models, other mechanisms such as loop extrusion have been advocated to be instrumental for maintaining a low degree of chromosomal entanglement~\citep{RackoStasiakPolymers2018,OrlandiniMarenduzzoMichielettoPNAS2019}.

To some inevitable extent though, physical entanglements are still expected to arise in eukaryotic chromosomes.

The recent work of Roca's lab  showed that knots {\it do} occur in eukaryotic minichromosomes {\em in vivo}, for instance during transcription, due to transient accumulation of entanglement~\citep{RocaNAR2017,ValdesNAR2019}.
On broader scales, various knots~\citep{VirnauPolymers2017}, and even links~\citep{SulkowskaSciRep2019}, were found in model mouse chromosomes obtained from single cell Hi-C~\citep{StevensNature2017}.
The genuineness of the entangled states was suggested by the systematic recurrence of certain knot types in independent instances of the reconstructed chromosomal structures~\citep{VirnauPolymers2017}.
These were obtained by imposing phenomenological constraints on an initially disconnected set of effective monomers, so we expect that a more defined knot spectrum could be obtained by using disentangled self-avoiding chains as the reference model.

\section{Conclusions}

To conclude, we have discussed experimental evidence and general physical mechanisms based on polymer theory that consistently point to an unusually low degree of entanglement expected in long eukaryotic chromosomes.
Such property, which is arguably essential for the capability of chromosomes to reconfigure as needed at various stages of the cell cycle, appears important for genomic modelling too.

We argued, in fact,  that the structural modelling of long chromosomes can benefit, both for realism and computational efficiency, by starting off with disentangled self-avoiding chains, {\it e.g.} mitotic-like ones, because their plasticity makes it possible to accommodate a large number of phenomenological constraints in a physically-viable manner, {\it i.e.} without deformations involving intra- or inter-chain crossings.

The latter are, of course, possible in {\it in vivo} systems thanks to the action of topoisomerase enzymes. An important open question regards, in fact, the extent to which these active mechanisms are
involved in the shaping the overall intra- and inter-chromosome architecture. This point, we believe, can be significantly advanced in future studies with a tight synergy of experiments and models~\citep{goloborodko2016,JostRosaVaillantEveraersReview2017,ValdesNAR2019}.

\section*{Conflict of Interest Statement}
The authors declare that the research was conducted in the absence of any commercial or financial relationships that could be construed as a potential conflict of interest.

\section*{Author Contributions}
All authors listed have made a substantial, direct and intellectual contribution to the work, and approved it for publication.

\section*{Funding}
The authors acknowledge support from the Italian Ministry of Education, MIUR.





%

\end{document}